\documentstyle[12pt]{article}

\begin{document}

\newcommand{\beq}{\begin{equation}}
\newcommand{\eeq}{\end{equation}}

\def\S{\Sigma}
\begin{titlepage}

\begin{center}

\hfill hep-th/9803198
\vskip 1.8cm
{\Large{\bf BPS Spectrum and Nahm Duality \\
of Matrix Theory Compactifications}}

\vskip 1.5cm
Shijong Ryang

\vskip .8cm
{\em Department of Physics \\ Kyoto Prefectural University of Medicine \\
Taishogun, Kyoto 603 Japan}

{\tt ryang@koto.kpu-m.ac.jp}
\end{center}

\vskip 2.0cm

\begin{center} {\bf Abstract} \end{center}
In the Matrix theories compactified on the $d$ dimensional torus with
$d = 5, 6, 7$ which are described by the theories of D$d$-branes in
the type IIA or IIB theory on the $d$ dimensional dual torus we analyze
the BPS bound states of the background D$d$-brane with other objects
by taking the matrix theory limit. Through the Nahm duality transformation
the flux and the momentum multiplets for these BPS states are shown to be
associated with the black holes and the black strings respectively from
the viewpoint of the weakly coupled type IIA string theory compactified
on the $d-1$ dimensional torus.

\vskip 4cm
\noindent March, 1998

\end{titlepage}

Many important evidences have been gathered in favor of the Matrix theory
description of M-theory \cite{BFSS}. The identification of M-theory with
the strong coupling limit of the type IIA string theory is realized by the
large $N$ limit accompanied with the large eleventh radius in the Matrix
theory. Explicit constructions of Matrix theory compactifications have been
studied in order to obtain full understandings of the compactification
of M-theory on compact manifolds. The Matrix theory compactified on
$T^d$ is shown to be equivalent to the $d + 1$ dimensional $U(N)$
supersymmetric Yang-Mills theory (SYM) on a dual torus $\tilde{T}^d$
\cite{WT,GRT,FHRS}. The U duality group of Matrix
theory compactified on $T^3$ is
given by $SL(3,Z) \times SL(2,Z)$ where the $SL(3,Z)$ is the geometric
symmetry group of $T^3$ and the $SL(2,Z)$ is
the S duality of $3 + 1$ dimensional
SYM theory. For $d > 3$ the SYM theories are strongly coupled in the
ultraviolet region so that extra degrees of freedom need to be added to
describe consistent full theories. The SYM theories for $d > 3$ are replaced
by some more general theories on the world-volumes of extended objects in
the string theory in order to controle the theories in the ultraviolet
region. The Matrix theory on $T^4$ is described in terms of the $5$-branes
of M-theory and the Matrix theory on $T^5$ is given by the theory of
NS$5$-branes in the type IIA or IIB string theory \cite{MR,BRS,NS1,SS,DVV}.
The theory of NS$5$-branes decouples from the bulk space-time modes to
become a well-defined theory as a Matrix theory in the limit that the
eleven dimensional Planck length and the string coupling constant
are taken to zero with the string length fixed. The compactification on
$T^6$ is proposed by the formulation of the world-volume theory of
Kaluza-Klein (KK) monopoles in eleven dimensions, which become the
D$6$-branes in the type IIA string theory language \cite{HL,LMS,BK,OG}.
Based on the world-volume theory of NS$5$-branes or D$6$-branes
the U duality group for the Matrix theory on $T^d$ with $d = 5$ or $6$ is
investigated as a symmetry group for the sprctrum of BPS bound states
 constructed by a NS$5$-brane or a D$6$-brane background with some other
objects. The structure of U duality group $E_d$ in the Matrix theory
compactified on $T^d$ with $d\geq 3$ is also systematically studied from
the effective SYM theory by imposing
additional requirements about generalized Montone-Olive duality and
permutation \cite{EGKR}, where the BPS spectrum is
considered to be reliably determined by this low energy effective theory
instead of the consistent full theory.

Recently through an appropriate matrix theory limit the Matrix theory
compactified on $T^d \times S^1$ with very small radius of the eleventh
spatial circle has been constructed as a weakly coupled IIA string theory
and shown to be related with the discrete light cone quantization of
the Matrix theory with finite radius of the light-like circle \cite{AS,NS2}.
In the Matrix theory on $T^d \times S^1$ in the discrete light cone gauge
that is described by a $d + 1$ dimensional $U(N)$ SYM theory with finite $N$,
the U duality group $E_d$ is shown to be enhanced to $E_{d+1}$ by the
Nahm duality symmetry (N duality) \cite{BO,CH} which exchanges the rank $N$
of the gauge group with some of the electic or magnetic fluxes \cite{HV}.
In the picture of Matrix theory on $T^d \times S^1$ the N duality acts
by exchanging the longitudinal radius with the radius of a space-like circle
so that it is related with the Lorentz invariance of the Matrix theory.

We are so interested in compactifications down to four dimensions that
it is a fascinating problem to elucidate the U dualities for the Matrix
theory compactifications down to lower dimensions. We aim to elaborate the
U duality groups for the Matrix theories compactified on $T^d \times S^1$
with $d = 5, 6, 7$. Making use of the matrix theory limit in the theories of
D$d$-branes in the type IIA or IIB string theories we will analyze
the BPS bound states of a background D$d$-brane and some other objects and
investigate the structures of U duality groups acting on these BPS states.
On the other hand in the viewpoint of the toroidally compactified type
IIA string theory at weak coupling the BPS states were originally
constructed as extreme black holes  or extreme black strings and shown
to form the U duality multiplets \cite{HT}. We will show that these
two kinds of pictures for BPS states are identified through the N duality
transformation.

The Matrix theory compactified on $T^d \times S^1$, a rectangular torus with
circles of radii $L_i, i=1,\cdots,d$ and $R$ is described by a $d + 1$
dimensional SYM theory on $\tilde{T}^d$, a dual torus with circles of radii
$\S_i, i=1,\cdots,d$ whose gauge coupling  is $g_{ym}$.
The relationships between them are given by
\beq
\S_i = \frac{l_s^2}{L_i}, \hspace{1cm} g_{ym}^2 = g_sl_s^{-3}
\prod^d_{i=1}\frac{l_s^2}{L_i},
\label{rel}\eeq
where the IIA string coupling  and length $g_s, l_s$
and the radius of the longitudinal dimension $R$ are related
with the Planck length $l_p$ as
$R = g_s^{2/3}l_p = g_sl_s$. The theory of coincident D$d$-branes in the
type IIA or IIB string theory on $\tilde{T}^d$ is generated by applying  T
duality transformations on all the circles of $T^d$  to the Matrix theory
on $T^d \times S^1$ that is the theory of D$0$-branes in the type IIA
string theory on $T^d$. At low energies this sysmtem reduces to the
previous $d + 1$ dimensional SYM theory on $\tilde{T}^d$. Since the string
coupling  of it is written as
\beq
 G_s = g_s\prod_{i=1}^d \frac{l_s}{L_i}
\label{stc}\eeq
the gauge coupling in the SYM language is specified by $G_s$ as
$g_{ym}^2 = G_sl_s^{d-3}$.

We start to devote ourselves to the $d = 5$ case. The Matrix theory on
$T^5 \times S^1$ is described by a theory of coincident D$5$-branes in the
type IIB string theory on $\tilde{T}^5$ with coupling $G_s$ and string
length $l_s$. In order to study BPS states which will fit into the
representations of U duality group we write down the masses of several
BPS objects such as D$p$-branes with $p = 1, 3, 5$, NS$1$-brane, NS$5$-brane
and pp-wave
\begin{eqnarray}
M_{D1} = \frac{\S_i}{G_sl_s^2}, & M_{D3} = \displaystyle\frac{\S_i\S_j\S_k}
{G_sl_s^4}, & M_{D5} = \frac{V_5}{G_sl_s^6}, \nonumber \\
M_{NS1} = \frac{\S_i}{l_s^2}, & M_{NS5} = \displaystyle
\frac{V_5}{G_s^2l_s^6}, & M_{pp} = \frac{1}{\S_i},
\label{BPS}\end{eqnarray}
where $V_d = \prod_{i=1}^d \S_i$. Here we take the matrix
theory limit \cite{AS,NS2}
where $l_p$ is taken to zero with the ratio $l_s/\sqrt{l_p}$ fixed
and the radii $L_i$ are measured in Planck unit;
\begin{eqnarray}
l_p \rightarrow 0, &    \nonumber \\
\frac{L_i}{l_p} = constant, & \displaystyle\frac{l_s}{\sqrt{l_p}}
= constant.
\label{mtl}\end{eqnarray}
In this limit $\S_i$ are kept constant and $G_s$, which is proportional to
$l_p^{-1}$, becomes strong coupling, while the string coupling constant
in the type IIA string theory is small as $g_s \sim l_p^{3/2}$, which
leads to a very small eleventh radius as $R \sim l_p^2$. Hence the
D$5$-brane is most heavy as $M_{D5} \sim l_p^{-2}$ and the others are
separated into two groups. One consists of the basic BPS states with masses
$M_{NS1}, M_{D3}$ and $M_{NS5}$ which are proportional to $l_p^{-1}$, and
the other includes $M_{D1}$ and $M_{pp}$ which are kept constant in the
matrix theory limit. The former basic BPS objects, whose masses are denoted
by $M_a$, combine with the D$5$-brane respectively to form non-threshold
bound states with masses squared $M^2 = M_{D5}^2 + M_a^2$.
Subtracting the background we have finite Yang-Mills energies in the matrix
theory limit
\beq
E_{NS1}^i = \frac{g_{ym}^2\S_i^2}{2V_5}, \hspace{1cm} E_{D3}^{ij} =
\frac{V_5}{2g_{ym}^2(\S_i\S_j)^2}, \hspace{1cm} E_{NS5} =
\frac{V_5}{2g_{ym}^6}
\label{for}\eeq
with finite $g_{ym}$. For any $d$ it is noted from (\ref{rel}) that the gauge
coupling is kept fixed in the decoupling limit (\ref{mtl}), while
the string coupling behaves as $G_s \sim l_p^{(3-d)/2}$. Each of the latter
basic BPS objects together with the background D$5$-brane makes a
threshold bound state with zero binding energy. The corresponding Yang-Mills
energies are given by
\beq
E_{D1}^i = \frac{\S_i}{g_{ym}^2}, \hspace{1cm} E_{pp}^i = \frac{1}{\S_i}.
\label{lat}\eeq
The former bound states in (\ref{for}) transform under $SL(5,Z)$ as
$5, 10,$ and $1$ respectively so that they give the $16$ of $SO(5,5,Z)$,
the U duality group of the Matrix theory on $T^5 \times S^1$, while the
latter bound states in (\ref{lat}) transforming as $5, 5$ under $SL(5,Z)$
give the $10$ of $SO(5,5,Z)$.

Now we will perform the S duality transformation defined by
\begin{eqnarray}
\hat{\S_i} = \S_i, & \hat{G_s} = G_s^{-1}, & \hat{l_s} = l_sG_s^{1/2},
\label{sdu}\end{eqnarray}
whcih convert the theory of D$5$-branes at strong coupling into that
of NS$5$-branes at weak coupling in the type IIB string theory. The matrix
theory limit characterized by $G_s \sim l_p^{-1}$, (\ref{mtl})
is changed into $\hat{G_s} \rightarrow 0, \hat{\S_i} = constant$ and
$\hat{l_s} = constant$. Through (\ref{sdu}) the masses of BPS objects
in (\ref{BPS}) are respectively mapped to
\begin{eqnarray}
M_{NS1} = \frac{\hat{\S_i}}{\hat{l_s}^2}, & M_{D3} =
\displaystyle\frac{\hat{\S_i}\hat{\S_j}\hat{\S_k}}{\hat{G_s}\hat{l_s}^4}, &
M_{NS5} = \frac{\hat{V_5}}{\hat{G_s}^2 \hat{l_s}^6}, \nonumber \\
M_{D1} = \frac{\hat{\S_i}}{\hat{G_s}\hat{l_s}^2}, & M_{D5} =
\displaystyle\frac{\hat{V_5}}{\hat{G_s}\hat{l_s}^6}, &
M_{pp} = \frac{1}{\hat{\S_i}}
\label{msd}\end{eqnarray}
as expected where $\hat{V}_5 = \prod_{i=1}^5\hat{\S_i}$ and
 the NS$5$-brane is most massive as $\hat{G_s}^{-2}$, corresponding to
the previous D$5$-brane. It produces a non-threshold bound state with the
second massive D$1$, D$3$ or D$5$-brane whose mass is proportional to
$\hat{G_s}^{-1}$. The finite Yang-Mills energies for them are constructed
in Ref. \cite{NS1}. On the other hand the more light NS$1$-brane or
pp-wave combines with the background NS$5$-brane
into a threshold bound state.
From this theory of NS$5$-branes in the type IIB string theory in the
limit that the string coupling vanishes with the string length kept fixed
the non-critical string theory with $(1, 1)$ supersymmetry is generated
and its low energy theory is further described by the six dimensional
SYM theory \cite{NS1,SS}. Moreover let us make the T duality transformation
about one of the five directions defined by $G_s^A =
\hat{G_s}\hat{l_s}/\hat{\S_i},  \S_i^A = \hat{l_s}^2/\hat{\S_i},
\S_j^A = \hat{\S_j} \; (j\neq i)$ and $l_{sA} = \hat{l_s}$.
The masses in (\ref{msd}) are well arranged into
\begin{eqnarray}
M_{pp} = \frac{1}{\S_i^A}, & M_{D2} =
\displaystyle\frac{\S_i^A \S_j^A}{G_s^A l_{sA}^3}, & M_{NS5} =
\frac{V_5^A}{(G_s^A)^2l_{sA}^6}, \nonumber \\
M_{D0} = \frac{1}{G_s^Al_{sA}}, & M_{D4} =
\displaystyle\frac{V_5^A}{G_s^Al_{sA}^5\S_i^A}, &  M_{NS1} =
\frac{\S_i^A}{l_{sA}^2}
\end{eqnarray}
with $V_5^A = \prod_{i=1}^5\S_i^A$.
Indeed if we take the $1$-st direction as $i$, one of the $M_{D1},
\hat{\S_1}/\hat{G_s}\hat{l_s}^2$ yields one $M_{D0}$ and the other four
$\hat{\S_j}/\hat{G_s}\hat{l_s}^2$ combine with the six $M_{D3} =
\hat{\S_1}\hat{\S_j}\hat{\S_k}/\hat{G_s}\hat{l_s}^4$ into the ten
$M_{D2}$. The remaining four $M_{D3} = \hat{\S_j}\hat{\S_k}\hat{\S_l}
/\hat{G_s}\hat{l_s}^4$ and one $M_{D5}$ become the five $M_{D4}$.
One $M_{NS5}$ turns to be again the mass of NS$5$-brane in the type
IIA string theory. The four $\hat{\S_j}/\hat{l_s}^2$ of the NS$1$-brane
with one $1/\hat{\S_1}$ of the pp-wave and the four $1/\hat{\S_j}$ of
the pp-wave with one $\hat{\S_1}/\hat{l_s}^2$ of the NS$1$-brane lead to
to the five $M_{NS1}$ and the five $M_{pp}$ respectively.
 In the matrix theory limit which is also specified
by $G_s^A \rightarrow 0, l_s^A = constant$ the heaviest NS$5$-brane
background combines with the lightest NS$1$-brane or pp-wave to form a
threshold bound state. This theory of NS$5$-branes in the type IIA string
theory becomes the $(2,0)$ string theory in the limit that the string
coupling vanishes with the string length held fixed. The succeeding zero
slope limit $l_s^A \rightarrow 0$ yields its low energy theory, that is
the $(2,0)$ field theory \cite{NS1,SS}.

Here we return to the theory of D$5$-branes in the type IIB string theory on
$\tilde{T}^5$. As discussed above the BPS bound states in this theory
are characterized by the basic constituent BPS objects. In the list of
BPS spectrum in (\ref{BPS}) we separate the fifth direction of the dual
five-torus from the other directions and classify the masses of the basic
BPS states as Table $1$.
\begin{table}
\begin{center}
\begin{tabular}{|c|c|c|c|c|} \hline
NS1 & $\frac{\S_i}{l_s^2} \; (4), \; \frac{\S_5}{l_s^2} \; (1)$ &  & 5 & \\
\cline{1-4}
D3 & $\frac{\S_i\S_j\S_5}{G_sl_s^4} \;(6)$ & $\frac{\S_i\S_j\S_k}{G_sl_s^4}
\; (4)$ &  10 & 16  \\  \cline{1-4}
NS5 &  & $\frac{V_5}{G_s^2l_s^6} \; (1)$ & 1 &      \\ \hline
D1 & $\frac{\S_5}{G_sl_s^2} \; (1)$ & $\frac{\S_i}{G_sl_s^2} \;(4)$ & 5 &
\\  \cline{1-4}
pp & $\frac{1}{\S_i} \; (4)$ & $\frac{1}{\S_5} \; (1)$ & 5 & 10  \\ \hline
\end{tabular}
\caption{The BPS states for the theory of D$5$-branes in the type IIB
string theory on $\tilde{T}^5$.}
\end{center}
\end{table}
In this table $ i = 1,\cdots, 4$ and for example with respect to the
D$3$-brane the $10$ of $SL(5,Z)$ decompose as $6 + 4$ under $SL(4,Z)$.
The Eqs. (\ref{rel}) and (\ref{stc}) make the basic BPS mass spectrum
change  into that for the theory of D$0$-branes in the type
IIA string theory on $T^5$, which is shown in  Table $2$ with
$V_d^L = \prod_{i=1}^dL_i$.
\begin{table}
\begin{center}
\begin{tabular}{|c|c|c|c|} \hline
pp & $\frac{1}{L_i}, \; \frac{1}{L_5}$ &  &  \\ \cline{1-3}
D2 & $\frac{L_iL_j}{g_sl_s^3}$ & $\frac{L_iL_5}{g_sl_s^3}$ & 16 \\
\cline{1-3}
NS5 &  & $\frac{V_5^L}{g_s^2l_s^6}$ &  \\ \hline
D4 & $\frac{V_5^L}{g_sl_s^5L_5}$ & $\frac{V_5^L}{g_sl_s^5L_i}$ &  \\
\cline{1-3}
NS1 & $\frac{L_i}{l_s^2}$ & $\frac{L_5}{l_s^2}$ & 10   \\ \hline
\end{tabular}
\caption{The BPS states for the theory of D$0$-branes in the type IIA
string theory of $T^5$.}
\end{center}
\end{table}
In the language of SYM theory on $\tilde{T^d}$ the N duality transformation
is defined by
\begin{eqnarray}
g_{ym}^2 \rightarrow g_{ym}^2\gamma_N^{d-4}, & \S_i \rightarrow
{\S_i}{\gamma_N}  \hspace{0.5cm} for \; i \neq d, \nonumber \\
\S_d \rightarrow \S_d, & l_s^2 \rightarrow {l_s^2}{\gamma_N},
\end{eqnarray}
where $\gamma_N = R/L_d = g_{ym}^2l_s^2\S_d/V_d$ \cite{BO}.
In Table $1$ under the N duality the $\S_5/G_sl_s^2 = \S_5/g_{ym}^2$
for the D$1$-brane and the $1/\S_i$ for the pp-wave are transformed into
$V_5/g_{ym}^4l_s^2 = V_5/G_s^2l_s^6$ for the NS$5$-brane and
$\S_i\S_j\S_k/g_{ym}^2l_s^2 = \S_i\S_j\S_k/G_sl_s^4$ for the D$3$-brane.
The $\S_5/l_s^2$ for the NS$1$-brane is also transformed into
$V_5/g_{ym}^2l_s^4$ which turns out to be $V_5/G_sl_s^6$, the mass of
D$5$-brane. The corresponding $1/L_5$ for the pp-wave in Table $2$ is
accordingly transformed into $1/R = 1/g_sl_s$ that is the D$0$-brane mass
in the IIA string theory on $T^5$. The others are not changed under the
N duality. In the viewpoint of the six dimensional weakly coupled type IIA
string theory compactified on $T^4$ the BPS states obtained by applying
the N duality transformation to the $16$ in Table $2$ are interpreted to
represent the following sixteen electric black holes; $4$ black holes
arising from pp-waves travelling in each of the $4$ toroidal dimensions,
one black hole given by the D$0$-brane, $6$ black holes arising from the
D$2$-branes wrapped around the $2$-torus, one black hole provided by the
D$4$-brane wrapped around the $4$-torus and $4$ black holes originating
in the NS$1$-branes wrapped around the $1$-torus. The BPS states transformed
from the $10$ consist of the following ten electric and magnetic black
strings; $4$ electric strings arising from the D$2$-branes wrapped around the
$1$-torus, one magnetic string generated from the NS$5$-brane wrapped around
the $4$-torus, $4$ magnetic strings arising from the D$4$-branes wrapped
around the $3$-torus and one electric string given by the unwrapped
NS$1$-brane.

Now we consider compactifying the Matrix theory on a six-torus down to
five dimensions. It is described by a theory of coincident D$6$-branes in the
type IIA string theory on $\tilde{T}^6$ with strong string coupling as
$G_s \sim l_p^{-3/2}$. The masses of BPS objects are given by
\begin{eqnarray*}
M_{D2} = \frac{\S_i\S_j}{G_sl_s^3}, & M_{D4} =
\displaystyle\frac{\S_i\S_j\S_k\S_l}{G_sl_s^5}, & M_{D6} = \frac{V_6}
{G_sl_s^7},
\end{eqnarray*}
\beq
M_{NS1} = \frac{\S_i}{l_s^2}, \hspace{0.5cm} M_{NS5} =
\displaystyle\frac{V_6}{G_s^2l_s^6\S_i},   \hspace{0.5cm}
 M_{pp} = \frac{1}{\S_i},  \hspace{0.5cm}
M_{KK} = \frac{V_6\S_i}{G_s^2l_s^8},
\eeq
where $M_{KK}$ is the mass of KK monopole  wrapped around the
six directions, where Taub-NUT direction is specified by $i$ \cite{HL,BK}.
The matrix theory limit (\ref{mtl}) shows that the D$6$-brane is most heavy
as $M_{D6} \sim l_p^{-2}$ and the $M_{NS1}, M_{D4}, M_{KK}$ behave as
$l_p^{-1}$ and the $M_{D2}, M_{pp}, M_{NS5}$ are finite. From these behaviors
it is noted that the background D$6$-brane can make non-threshold bound
states with the NS$1$-brane, D$4$-brane and KK monopole respectively, whose
finite Yang-Mills energies are
\begin{eqnarray}
E_{NS1}^i = \frac{g_{ym}^2\S_i^2}{2V_6}, & E_{D4}^{ij} =
\displaystyle\frac{V_6}{2g_{ym}^2(\S_i\S_j)^2}, & E_{KK}^i =
\frac{V_6\S_i^2}{2g_{ym}^6}.
\end{eqnarray}
They build up a flux multiplet which is a $27$ of $E_6(Z)$, the U duality
group for the Matrix theory compactified on $T^6 \times S^1$.
The $27$ decompose as $6 + 15 + 6$ under $SL(6,Z)$. On the other hand the
D$2$-brane, pp-wave and NS$5$-brane can respectively combine with the
background D$6$-brane to make threshold bound states whose finite
Yang-Mills energies are
\begin{eqnarray}
E_{D2}^{ij} = \frac{\S_i\S_j}{g_{ym}^2}, & E_{pp}^i =
\displaystyle\frac{1}{\S_i}, & E_{NS5}^i = \frac{V_6}{g_{ym}^4\S_i}.
\end{eqnarray}
These BPS bound states transform as the $15, 6, 6$ under $SL(6,Z)$ and form
a momentum multiplet represented by a $\overline{27}$ of $E_6(Z)$.

Let us single out the sixth direction of the dual six-torus from the others
in order to classify the masses of the basic BPS states for the theory of
D$6$-branes at strong coupling in the IIA string theory on $\tilde{T}^6$
in Table $3$.
\begin{table}
\begin{center}
\begin{tabular}{|c|c|c|c|c|} \hline
NS1 & $\frac{\S_i}{l_s^2} \; (5), \; \frac{\S_6}{l_s^2} \; (1)$ &  & 6 &  \\
\cline{1-4}
D4 & $\frac{V_6}{G_sl_s^5\S_i\S_j} \; (10)$ & $\frac{V_6}{G_sl_s^5\S_i\S_6}
\; (5)$ & 15 & 27 \\ \cline{1-4}
KK & $\frac{V_6\S_6}{G_s^2l_s^8} \; (1)$ & $\frac{V_6\S_i}{G_s^2l_s^8} \;
(5)$ & 6 &  \\ \hline
NS5 & & $\frac{V_6}{G_s^2l_s^6\S_i} \; (5), \; \frac{V_6}{G_s^2l_s^6\S_6}
\; (1)$ & 6 &  \\ \cline{1-4}
D2 & $\frac{\S_i\S_6}{G_sl_s^3} \; (5)$ & $\frac{\S_i\S_j}{G_sl_s^3} \;(10)$
& 15 & $\overline{27}$  \\ \cline{1-4}
pp & $\frac{1}{\S_i} \; (5)$ & $\frac{1}{\S_6} \; (1)$ & 6 &  \\ \hline
\end{tabular}
\caption{The BPS states for the theory of D$6$-branes in the type IIA
string theory on $\tilde{T}^6$.}
\end{center}
\end{table}
In this table the BPS states are distributed in a similar way to Table $1$.
Through (\ref{rel}) and (\ref{stc}) the BPS states in Table $3$ are
transformed back into those for the theory of D$0$-branes at weak coupling
in the type IIA string theory on $T^6$ as Table $4$.
\begin{table}
\begin{center}
\begin{tabular}{|c|c|c|c|} \hline
pp & $\frac{1}{L_i}, \; \frac{1}{L_6}$ &  &  \\ \cline{1-3}
D2 & $\frac{L_iL_j}{g_sl_s^3}$ & $\frac{L_iL_6}{g_sl_s^3}$ & 27 \\
\cline{1-3}
NS5 & $\frac{V_6^L}{g_s^2l_s^6L_6}$ & $\frac{V_6^L}{g_s^2l_s^6L_i}$ &
\\ \hline
KK &  & $\frac{V_6^LL_i}{g_s^2l_s^8}, \; \frac{V_6^LL_6}{g_s^2l_s^8}$ &
\\ \cline{1-3}
D4 & $\frac{V_6^L}{g_sl_s^5L_iL_6}$ & $\frac{V_6^L}{g_sl_s^5L_iL_j}$ &
$\overline{27}$ \\ \cline{1-3}
NS1 & $\frac{L_i}{l_s^2}$ & $\frac{L_6}{l_s^2}$ &  \\ \hline
\end{tabular}
\caption{The BPS states for the theory of D$0$-branes in the type IIA string
theory on $T^6$.}
\end{center}
\end{table}
In this stage we will make the N duality transformation. The $\S_6/l_s^2$
of NS$1$ and the $V_6/G_s^2l_s^6\S_6$ of NS$5$ in Table $3$ are
transformed into the masses of D$6$-brane and D$0$-brane, $V_6/G_sl_s^7$
and $1/G_sl_s$ respectively, which further correspond to the masses of
D$0$-brane and D$6$-brane for the theory of D$0$-branes in the IIA
string theory on $T^6, 1/g_sl_s$ and $V_6^L/g_sl_s^7$. The
$\S_i\S_6/G_sl_s^3$ of D$2$ and the $1/\S_i$ of pp are interchanged by
the $V_6\S_i/G_s^2l_s^8$ of KK and the $V_6/G_sl_s^5\S_i\S_6$ of D$4$
respectively. The others are invariant under the N duality transformation.
From Table $4$ we deduce that the flux multiplet $27$ is transformed
under the N duality into the twenty seven electric black holes.
Indeed these states are interpreted in the viewpoint of the type IIA string
theory compactified on $T^5$ at weak coupling as follows;
$6$ black holes provided by the pp-waves travelling in each of the $5$
toroidal dimensions and the one D$0$-brane, 15 black holes generated by the
$10$ D$2$-branes wrapped around the $2$-torus and the $5$ NS$1$-branes
wrapped around the $1$-torus, and $6$ black holes given by the one
NS$5$-brane wrapped around the $5$-torus and the $5$ D$4$-branes wrapped
around the $4$-torus. It is interesting to note that the momentum multiplet
$\overline{27}$ is transformed into the twenty seven magnetic black strings;
$6$ black strings generated by the $5$ KK magnetic monopoles, that are
Taub-NUT $5$-branes, wrapped around the $5$-torus and the one D$6$-brane
wrapped around the $5$-torus, $15$ black strings produced by the $5$
NS$5$-branes wrapped around the $4$-torus and the $10$ D$4$-branes wrapped
around the $3$-torus, and $6$ black strings given by the $5$ D$2$-branes
wrapped around the $1$-torus and the one unwrapped NS$1$-brane.
In this way we observe that these N duality transformed states $27,
\overline{27}$ are just extreme electric black holes and extreme magnetic
black strings which explicitly appear in the weakly coupled type IIA
string theory compactified on $T^5$ as argued in Ref. \cite{HT}.
The BPS electric black holes $27$ decompose as $16 + 10 + 1$ under
$SO(5,5,Z)$, where the $1$ corresponds to the one
NS$5$-brane wrapped around the
$5$-torus and the $16$ are associated with the D$0$, D$2$, D$4$-branes
and the $10$ correspond to the five directions of pp-wave and the five
directions of NS$1$-brane winding.

Here we extend the previous analysis to consider the Matrix theory
compactified on $T^7 \times S^1$ that is decribed by a theory of coincident
D$7$-branes in the type IIB string theory on $\tilde{T^7}$. The BPS
states are specified by the following masses;
\begin{eqnarray}
M_{D3} = \frac{\S_i\S_j\S_k}{G_sl_s^4}, & M_{D5} = \displaystyle
\frac{\S_{i_1} \cdots \S_{i_5}}{G_sl_s^6}, & M_{D7} = \frac{V_7}
{G_sl_s^8}, \nonumber \\
M_{NS1} = \frac{\S_i}{l_s^2}, & M_{pp} = \displaystyle\frac
{1}{\S_i}, & M_{KK} = \frac{V_7\S_i}{G_s^2l_s^8\S_j}
\end{eqnarray}
and
\begin{eqnarray}
M_{(NS5)} = \frac{V_7\S_i\S_j}{G_s^2l_s^{10}}, &
M_{(D1)} = \displaystyle\frac{V_7^2}{G_s^3l_s^{14}\S_i},
\end{eqnarray}
which are masses for the new type of states that we represent by $(NS5)$
and $(D1)$. There are further basic BPS states whose masses are
$V_7^2/G_s^3l_s^{12}\S_i\S_j\S_k$ and $V_7^2\S_i/G_s^4l_s^{16}$, that will
not be analyzed. From $M_{NS5} = \S_{i_1} \cdots \S_{i_5}/G_s^2l_s^6$ the
mass $M_{(NS5)}$ is produced by making two T duality transformations along
two directions transverse to the world-volume of the NS$5$-brane and
the $M_{(D1)}$ is also interpreted to be associated with the rolled-up
Taub-NUT $6$-brane in Ref. \cite{BO}.
The $M_{KK}$ is regarded as the mass of
Taub-NUT $5$-brane with the $i$-th Taub-NUT direction, which is wrapped
around the $6$-torus transverse to the $j$-th direction. Analyzing the
behaviors of masses in the matrix limit we get the structure of the BPS
bound states. The most heavy D$7$-brane with $M_{D7} \sim l_p^{-2}$
can combine with the NS$1$-brane, D$5$-brane, (D$1$)-brane and
(NS$5$)-brane respectively whose masses behave as $l_p^{-1}$, to construct
non-threshold bound states that form a flux multiplet, $56$ of $E_7(Z)$,
the U duality group for the Matrix theory on $T^7 \times S^1$.
The $M_{KK}, M_{D3}$ and $M_{pp}$ show no $l_p$ dependence so that
the KK monopole, D$3$-brane and pp-wave combine with the background
D$7$-brane to build up threshold bound states that belong to a momentum
multiplet, while the D$1$-brane and NS$5$-brane cannot build up
bound states with finite non-zero energies since the $M_{D1} =
\S_i/G_sl_s^2$ and $M_{NS5}$ are proportional to $l_p$.
The finite Yang-Mills energies of the bound states for the flux multiplet
are derived as
\beq
E_{NS1}^i = \frac{g_{ym}^2\S_i^2}{2V_7}, \hspace{0.5cm} E_{D5}^{ij} =
\displaystyle\frac{V_7}{2g_{ym}^2(\S_i\S_j)^2}, \hspace{0.5cm} E_{(D1)}^i =
\frac{V_7^3}{2g_{ym}^{10}\S_i^2}, \hspace{0.5cm} E_{(NS5)}^{ij} =
\frac{V_7\S_i^2\S_j^2}{2g_{ym}^6},
\eeq
which transform as the $7, 21, 7$ and $21$ under $SL(7,Z)$.

Separating the seventh direction of the dual seven-torus from the others
we carry out the classification of masses of basic BPS states for the theory
of D$7$-branes at strong coupling in the type IIB string theory on
$\tilde{T}^7$ in Table $5$.
\begin{table}
\begin{center}
\begin{tabular}{|c|c|c|c|c|} \hline
NS1 & $\frac{\S_i}{l_s^2} \; (6), \; \frac{\S_7}{l_s^2}  \; (1)$ &  & 7 & \\
\cline{1-4}
D5 & $\frac{\S_i\S_j\S_k\S_l\S_7}{G_sl_s^6} \; (15)$ &
$\frac{\S_i\S_j\S_k\S_l\S_m}{G_sl_s^6} \; (6)$ & 21 &  \\ \cline{1-4}
(D1) & $\frac{V_7^2}{G_s^3l_s^{14}\S_7} \;(1)$ & $\frac{V_7^2}
{G_s^3l_s^{14}\S_i} \; (6)$ & 7 & 56 \\ \cline{1-4}
(NS5) & $\frac{V_7\S_i\S_7}{G_s^2l_s^{10}} \; (6)$ & $\frac{V_7\S_i\S_j}
{G_s^2l_s^{10}} \; (15)$ & 21 &  \\ \hline
KK & $\frac{V_7\S_7}{G_s^2l_s^8\S_i} \; (6)$ & $\frac{V_7\S_i}
{G_s^2l_s^8\S_7} \; (6), \; \frac{V_7\S_i}{G_s^2l_s^8\S_j} \; (30)$ &
42 &  \\ \cline{1-4}
D3 & $\frac{\S_i\S_j\S_7}{G_sl_s^4} \; (15)$ & $\frac{\S_i\S_j\S_k}
{G_sl_s^4} \; (20)$ & 35 &  \\ \cline{1-4}
pp & $\frac{1}{\S_i} \; (6)$ & $\frac{1}{\S_7} \; (1)$ & 7 &  \\ \hline
\end{tabular}
\caption{The BPS states for the theory of D$7$-branes in the type IIB string
theory on $\tilde{T}^7$.}
\end{center}
\end{table}
The distribution of the BPS states in this table is constructed by regarding
those in Tables $1, 3$ as suggestive guides, however there are some
different structures. Comparing the BPS spectrums of the type
IIB theories on $\tilde{T}^5, \tilde{T}^6, \tilde{T}^7$
in Tables $1, 3, 5$ we note that as the dimensions of dual tori increase
the NS$5$-brane in the flux multiplet in Table $1$ turns to be in the
momentum multiplet as shown in Table $3$, which is also the case for the KK
monopole in the flux multiplet in Tabel $3$ that
goes into the momentum multiplet
in Table $5$. These basic BPS states are transformed back into those for the
theory of D$0$-branes at weak coupling in the type IIA string theory on
$T^7$ as Table $6$.
\begin{table}
\begin{center}
\begin{tabular}{|c|c|c|c|} \hline
pp & $\frac{1}{L_i}, \frac{1}{L_7}$ &  &  \\ \cline{1-3}
D2 & $\frac{L_iL_j}{g_sl_s^3}$ & $\frac{L_iL_7}{g_sl_s^3}$ &  \\
\cline{1-3}
$(D1)^*$ & $\frac{V_7^LL_7}{g_s^3l_s^9}$ & $\frac{V_7^LL_i}{g_s^3l_s^9}$ &
56 \\ \cline{1-3}
$(NS5)^*$ & $\frac{L_iL_jL_kL_lL_m}{g_s^2l_s^6}$ & $\frac{V_7^L}
{g_s^2l_s^6L_iL_j}$ &  \\ \hline
KK & $\frac{V_7^LL_i}{g_s^2l_s^8L_7}$ & $\frac{V_7^LL_7}{g_s^2l_s^8L_i},
\; \frac{V_7^LL_i}{g_s^2l_s^8L_j}$ &  \\ \cline{1-3}
D4 & $\frac{L_iL_jL_kL_l}{g_sl_s^5}$ & $\frac{L_iL_jL_kL_7}{g_sl_s^5}$
&  \\ \cline{1-3}
NS1 & $\frac{L_i}{l_s}$ & $\frac{L_7}{l_s}$ &  \\ \hline
\end{tabular}
\caption{The BPS states for the theory of D$0$-branes in the type IIA string
theory on $T^7$.}
\end{center}
\end{table}
Under the N duality transformation in the type IIB string theory on
$\tilde{T}^7$ the $\S_7/l_s^2$ of NS$1$, the $V_7^2/G_s^3l_s^{14}\S_7$
of (D$1$) and the $V_7\S_i/G_s^2l_s^8\S_7$ of KK in Table $5$ are
mapped to the masses of D$7$-brane or D$1$-brane, $V_7/G_sl_s^8,
\S_7/G_sl_s^2, \S_i/G_sl_s^2$, which further turn out to be those of
D$0$-brane or D$6$-brane, $1/g_sl_s, V_7^L/g_sl_s^7L_7, V_7^L/g_sl_s^7L_i$
respectively for the theory of D$0$-branes in the type IIA string theory
on $T^7$. The $V_7\S_7/G_s^2l_s^8\S_i$ of KK, the $\S_i\S_j\S_7/
G_sl_s^4$ of D$3$ and the $1/\S_i$ of pp are mapped to $V_7^2
/G_s^3l_s^{14}\S_i$ of (D$1$), $V_7\S_i\S_j/G_s^2l_s^{10}$ of (NS$5$) and
$\S_i\S_j\S_k\S_l\S_m/G_sl_s^6$ of D$5$ to each other. The other states are
invariant under the N duality transformation. Gathering these results we
find that the N duality transformation maps the flux multiplet $56$ in
Table $6$ to the following electric and magnetic black holes for the
weakly coupled type IIA string theory compactified on $T^6$ \cite{HT};
$7$ electric black holes originating in the $6$ pp-waves travelling in each
of the $6$ toroidal dimensions and one D$0$-brane, $21$ electric black holes
provided by the $15$ D$2$-branes wrapped around the $2$-torus and the $6$
NS$1$-branes wrapped around the $1$-torus, $7$ magnetic black holes arising
from the one D$6$-brane wrapped around the $6$-torus and the $6$ KK
monopoles wrapped around the $6$-torus with a Taub-NUT direction in each of
the $6$ toroidal dimensions, $21$ magnetic black holes given by the $6$
NS$5$-branes wrapped around the $5$-torus and the $15$ D$4$-branes wrapped
around the $4$-torus.

In conclusion we have carried out the N duality transformations concretely
for the masses of basic BPS states and found the relations between the BPS
spectrums of the Matrix theory compactifications and those of the weakly
coupled type IIA string theory compactifications. The flux multiplets in the
Matrix theories compactified on $T^d \times S^1$ with $d = 5, 6, 7$ are at
first sight interpreted as the black holes as well as the black strings
from the viewpoint of the weakly coupled type IIA string theories
compactified on $T^{d-1}$. Under the N duality transformation, however
they are mapped just only to the extremal BPS black holes.
Analogously the momentum multiplets are transformed to the extremal BPS
black strings. We have observed that for $d = 5, 6$ the flux multiplets
$16$ of $SO(5,5,Z), 27$ of $E_6(Z)$ correspond to the electric black holes
and the momentum multiplets $10, \overline{27}$ to the $5$ magnetic and $5$
electric black strings, the twenty seven magnetic black strings, while
for $d =7$ the flux multiplet $56$ of $E_7(Z)$ turns out to be identified
with the $28$ electric and $28$ magnetic black holes, thus providing
further evidence for the Matrix theory compactifications. The matrix
theory limit proves useful when dividing the BPS states into the flux
and the momentum multiplets. By interchanging the original role of the
eleventh direction with that of one of the other directions we can naturally
incorporate the BPS black holes or black strings of the weakly coupled type
IIA string theory into the Matrix theory description. It would be interesting
to make clear the rich structures of the BPS spectrums in the
compactifications of various kinds of the ten dimensional string theories
in terms of the Matrix theory description. The N duality transformation
would play an important role for this elucidation and in particular
give a clue for relations between the strongly coupled  IIA  theory
compactifications and the weakly coupled IIA theory compactifications.

\end{document}